\pdfoutput=1
\documentclass[prd,preprint,floats,floatfix,aps,nofootinbib,preprintnumbers]{revtex4}
\usepackage{graphics,graphicx,psfrag}
\usepackage{amsmath,amssymb}
\usepackage[sort&compress]{natbib}
\usepackage{bm}	% for boldface in math
\usepackage{verbatim}
\usepackage{multirow}
\usepackage{subfigure}
\usepackage{ulem}
\usepackage{color}
\usepackage[linktocpage=true,bookmarksnumbered=true,colorlinks=true,plainpages,linkcolor=blue,citecolor=red]{hyperref}
\usepackage[utf8]{inputenc}  %per INSPIRE directions

\graphicspath{{./figs_v3/}}

\def\issue(#1,#2,#3){{\bf #1}, #2 (#3)}

\catcode`\@=11
\def\lsim{\mathrel{\mathpalette\@versim<}}
\def\gsim{\mathrel{\mathpalette\@versim>}}
\def\@versim#1#2{\vcenter{\offinterlineskip
\ialign{$\m@th#1\hfil##\hfil$\crcr#2\crcr\sim\crcr } }}
\catcode`\@=12

\catcode`@=12

\newcommand{\newc}{\newcommand}
\newc{\wt}{\widetilde}
\newc{\ra}{\rightarrow}
\usepackage{hyperref}
\def\beq {\begin{equation}}
\def\eeq {\end{equation}}
\def\bi {\begin{itemize}}
\def\ei {\end{itemize}}
\def\bea {\begin{eqnarray}}
\def\eea {\end{eqnarray}}

\def\Zp{Z^{\prime}}

\def\issue(#1,#2,#3){{\bf #1}, #2 (#3)}

\hypersetup{
   colorlinks=true,       % false: boxed links; true: colored links
   linkcolor=blue,        % color of internal links
   citecolor=red,         % color of links to bibliography
   filecolor=magenta,      % color of file links
   }

\begin{document}

%\preprint{
%{\vbox {
%\hbox{\bf MSUHEP-}
%\hbox{\today}
%}}}
%\vspace*{2cm}

%
\title{$R_{K}$ anomalies and simplified limits on $\Zp$ models at the LHC}
\vspace*{0.25in}   % makes space between address and abstract

\author{R. Sekhar Chivukula}
\email{sekhar@msu.edu}
\author{Joshua Isaacson}
\email{isaacs21@msu.edu}
\author{Kirtimaan A. Mohan}
\email{kamohan@msu.edu}
\author{Dipan Sengupta}
\email{dipan@msu.edu}
\author{Elizabeth H. Simmons}
\email{esimmons@msu.edu}
\affiliation{\vspace*{0.1in}
 Department of Physics and Astronomy\\
Michigan State University \\
 567 Wilson Road, East Lansing U.S.A.\\
}
%\vspace*{0.25 in} % makes space between address and abstrac

\begin{abstract}{  
The LHCb collaboration has recently reported  a 2.5 $\sigma$ discrepancy with respect to the predicted
value in a test of lepton universality in the ratio  $R_{K^*}= \hbox{BR}(B \to K^* \mu^+ \mu^-) / \hbox{BR}(B \to K^* e^+ e^-)$. 
Coupled with an earlier observation of  a similar anomaly in $R_{K}$,  this has generated significant  excitement.  A number of new physics scenarios 
have been proposed to explain the anomaly.
 In this work we consider simplified limits on $\Zp$ models from ATLAS and CMS searches for new resonances in dilepton and dijet modes, and we use the simplified limits variable $\zeta$ to correlate the results of the resonance and B-decay experiments. By examining minimal $Z'$ models that can accomodate the observed LHCb results, we show that the high-mass resonance search results are begining to be sensitive to these models and that future results will be more informative.}
\end{abstract}
\maketitle

\section{Introduction}

Run-2 of the LHC is well under away and the hunt for new physics has gathered pace. While no
clear signature of physics beyond the standard model (BSM) has been seen yet in the CMS or ATLAS experiments, 
a large swath of parameter space has been explored in the context of various models. A complementary strategy is explored by the 
LHCb collaboration (in addition to the experiments Belle, BaBar and KEK),  where deviations in $B$  physics observables from predicted standard model values  could potentially be a signature of new physics.  Over the years,  several $B$-physics processes have shown deviations from Standard 
model predictions \cite{1412.7515}. While there remain certain caveats about state of the art SM calculations, specially in estimating higher order QCD contributions, it is worthwhile to consider new particles that can explain the anomalies and to explore the potential consequences of these particles in other experiments. 

Very recently the LHCb experiment has observed an anomaly in a test of lepton flavor universality in the decay $B\to K^{*} \ell^{+}\ell^{-}$ 
with a statistical significance of $\rm 2.5~\sigma$ \cite{RKStar}. Coupled with an earlier observation in the decay  $B\to K \ell^{+}\ell^{-}$, 
that had a similar anomaly \cite{arXiv:1406.6482}, this has generated a significant amount of excitement in the community.  A number of studies have already 
appeared in this context, most of which perform a global fit of the operators that contribute to this anomaly, and quantify 
the discrepancy in terms of deviations from the corresponding standard model values \cite{Altmannshofer:2017yso, Alok:2017jaf,Capdevila:2017bsm,1704.05438,1704.05446,1704.05447,1704.05444,1704.07397,1704.05672,Ghosh:2017ber,Bardhan:2017xcc}.  Some efforts have also been devoted to an explanation 
with additional gauge bosons ($Z'$), lepto-quarks, and light particles 
\cite{Belanger:2015nma,Kamenik:2017tnu,Chiang:2017hlj,GarciaGarcia:2016nvr,Altmannshofer:2016jzy,Allanach:2015gkd,Crivellin:2015mga,Crivellin:2015era,Crivellin:2016ejn,Glashow:2014iga,Bhattacharya:2014wla,Bhattacharya:2016mcc,1704.05835,1704.05849,1704.06200,1704.08158,1705.00915,1705.03447,1705.03465,1705.05643}.
We also note that similar  anomalies  have been observed in the measurement 
of $R_{D}$ and $R_{D^{*}}$ and still persist \cite{rd}. 

While low energy $B$-physics observables can provide indirect clues to the plausible nature  of new physics models, 
the new particles in these models will have to eventually be directly discovered.  Since we have not observed any new particles  at the LHC, any explanation of the anomaly has to be consistent with CMS and ATLAS search bounds on such objects. In this paper, we use the simplified  limits framework \cite{Chivukula:2016hvp}, to constrain the parameter space available to a general phenomenological $Z'$ model.  On the one hand, we employ this framework for its originally identified purpose of presenting collider data on resonance searches in a form that facilitates identification of production and decay channels that could explain a new excess. On the other, we also show how to use this framework to simultaneously express different pieces of theoretical and experimental information in a unified language that provides an overarching picture of the viable parameter space of the model.

 First, we  use the $R_{K}$ and $R_{K^{*}}$ anomalies to put an upper bound on the value of the ``simplified limits variable" $\zeta$; then, we apply theoretical considerations to obtain a lower bound on $\zeta$.  Having identified a swath of parameter space within which a $Z'$ model would be both theoretically self-consistent and able to explain the LHCb observables, we consider how CMS and ATLAS dijet and dilepton data further constrain $\zeta$. 
We show that high-mass LHC resonance search results are begining to be sensitive to the general class of $Z'$ models that could be responsible for the $R_{K}$ and $R_{K^{*}}$ anomalies -- and that future results will be more sensitive.

\section{Quantifying the anomaly}

As mentioned earlier, there are two  independent observations by LHCb that point to  lepton flavor violation. The first of them is the ratio \cite{Aaij:2014ora},
\begin{equation}
R_K =\frac{{\rm BR} \left ( B^+ \to K^+ \mu ^+ \mu ^-\right )}{{\rm BR} \left ( B^+ \to K^+ e^+ e^-\right )}= 0.745\pm 0.09_{\rm stat} \pm0.036_{\rm syst} \, ;
\label{eqn:RK}
\end{equation}
in the bin of $\rm q^{2}~\in~ [1,6]~ GeV^{2}$. The corresponding predicted SM value is   $R_{K} = 1.0004 (8)$ \cite{Hiller:2003js}. 
%The significance of this  deviation is about $2.6~\sigma$.  
The second measured anomaly is the recently measured value of $R_{K^{*}}$ 
 ~\cite{RKStar},
\begin{equation}
R_{K^*}=\frac{\hbox{BR}(B \to K^* \mu^+ \mu^-) }{ \hbox{BR}(B \to K^* e^+ e^-) }  \ .
\label{eqn:RK*}
\end{equation}
 The measurements  in two different dilepton invariant mass squared ($q^2$) bins using the $1\,\text{fb}^{-1}$ data set from LHCb~\cite{Aaij:2014ora} yield, 
 \begin{equation}
\rm R_{K^* [0.045,1.1] ~GeV^{2}}= 0.660^{+0.110}_{-0.070} \pm 0.024 , ~~ R_{K^* [1.1,6] ~GeV^{2}}= 0.685^{+0.113}_{-0.069} \pm 0.047 ~.
\label{eqn:RK*}
\end{equation}
The SM value corresponds to \cite{Hiller:2003js,1605.07633} \footnote{Note that the most up to date values of the SM prediction including QED effects, according to \cite{1605.07633} are: 
   $\rm R_{K~  [1,6]~ GeV^{2}} = 1.00 \pm 0.01$; while 
$ \rm R_{K^* [0.045,1.1] ~GeV^{2}}=  0.901 \pm 0.028 ,~~  R_{K^* [1.1,6] ~GeV^{2}}=  1.00 \pm 0.01. $ While 
this changes the fit values obtained in \cite{1704.05446} slightly, the features of this analysis do not change significantly. 
 }, 
 \begin{equation}
\rm R_{K^* [0.045,1.1] ~GeV^{2}}=  0.920(7) ,~~  R_{K^* [1.1,6] ~GeV^{2}}=  0.996 (2) ~.
\label{eqn:RK*}
\end{equation}
Individually, each of these measurements point to $\sim 2.5 \sigma$ deviations from the standard model predictions. \footnote{Note that the anomaly 
in the branching ratio is observed entirely in the muons. The branching ratio to electrons agrees with that of the SM prediction. } However, global fits indicate that deviations from the SM are found to be about $4\sigma$~\cite{1704.05438,1704.05446}. In the past, several analyses
have exploited angular distributions in $B\to K^{*}\mu^{+}\mu^{-}$ \cite{1308.1707,1512.04442,1612.05014} to claim that deviations from SM in global fits can be 
between 4-5~$\sigma$ \cite{1310.2478,1510.04239,1603.00865,1703.09189}. Yet  it has also been observed that hadronic uncertainties can be significant and 
can bring the significance down considerably \cite{hep-ph/0106067,hep-ph/0404250,0807.2589,1006.4945,1101.5118,1211.0234,1212.2263,1310.3887,1406.0566,1407.8526,1412.3183,
1503.05534,1701.08672,1702.02234}.

 The $b\to s ll$ transitions can be studied in the language of effective Lagrangians. To leading order in $G_F$ the effective Hamiltonian for these transitions,\footnote{Here and throughout it is understood that the Wilson coefficients and the corresponding matrix elements are renormalized at a scale of order $m_b$.} at low energy in SM, is:
\begin{equation}
\mathcal{H}^{SM}_{eff} = \frac{4	G_F}{\sqrt{2}} \sum_{p=u,c} V_{pb}V^{\star}_{ps}\left(
C_1 \mathcal{O}_{1}^{p} + C_{2} \mathcal{O}_{2}^{p}  + \sum_{i=3}^{10} C_{i}\mathcal{O}_{i}
\right)
\end{equation}
The operators that contribute to the effective Hamiltonian 
can be classified as \cite{1704.05446},
\begin{itemize}
	\item $\mathcal{O}_{1,2}$: Current-current operators.
	\item $\mathcal{O}_{3,4,5,6}$: QCD penguins.
\item $\mathcal{O}_7$ : Electromagnetic penguin.
\item $\mathcal{O}_{8}$: Chromo-magnetic operator.
\item $\mathcal{O}_{9,10}$: Semi-leptonic operators.
\end{itemize}
 
BSM effects can be studied by modifying Wilson coefficients $C_i$ and by supplementing the Lagrangian with chirally flipped versions of the operators $\mathcal{O}_i$.
It can be shown that the operators $\mathcal{O}_{1} ... \mathcal{O}_{8}$ do not 
contribute directly to lepton flavor violation. 
%The contribution to the decay width $\Gamma ( B\to K l^{+} l^{-} )$ comes from $C_{9}$ and $C_{10}$ operators.
 Out of all the 
operators described above, the four that can potentially explain the deficits in the measurements of  $R_K$ and $R_{K^{*}}$ are 
\begin{align}
\mathcal{O}_{9}^{(\prime)}=\frac{\alpha_{em}}{4 \pi}\left(\bar{s}\gamma^{\mu}P_{L(R)} b \right)\left(\bar{l}\gamma_{\mu}l\right), \quad
\mathcal{O}_{10}^{(\prime)}=\frac{\alpha_{em}}{4 \pi}\left(\bar{s}\gamma^{\mu}P_{L(R)} b \right)\left(\bar{l}\gamma_{\mu}\gamma_{5}l\right)~,
\end{align}
where the unprimed operators involve left-handed quark currents, the primed ones involve right-handed (chirality-flipped) currents, and $P_{L}, P_{R}$ are the left- and right-handed projection operators respectively.

We follow the analysis of \cite{1704.05446} to quantify 
the deviation in terms of non-universal BSM contributions ($\delta C_{9}, ~\delta C_{10}$). Although the deficit could arise from a combination of lepton flavor violating effects in both 
electron and muon sectors, for simplicty we assume that the new physics contribution is muon specific. For the $R_{K}$ anomaly, since the standard model contribution is $C_{9}^{SM}\simeq -C_{10}^{SM}= 4.27$, a muon specific BSM 
contribution requires, 
\begin{equation}
~\rm either~\delta C_{9}^{(')\mu} ~\simeq~ -1~,~~~or~ \delta C_{10}^{(')\mu} ~= ~+1 ~.
\end{equation}
Equivalently, one can express the above in terms of a leptonic left handed combination \cite{1704.05446}, 
 \begin{equation}
 \delta C_{9}^{\mu}~ =~ - \delta C_{10}^{\mu} ~= ~ -0.5 ~.
\end{equation}

The prediction for the decay width for $\Gamma ( B\to K^{*} l^{+} l^{-}  )$ is more involved. On decomposing the expression for width into a transverse and longitudinal part it can be seen that 
the longitudinal part of the width differs from   $\Gamma ( B\to K l^{+} l^{-}  )$
by a relative minus sign in the interference of the SM contribution and its chirally flipped 
counterpart. This change in sign implies that a simultaneous reduction in both $R_K$ and $R_{K^{*}}$
cannot be explained by the primed, chirally flipped, operators alone. For instance, a drop in $R_{K}$ via a negative 
contribution in the chirally flipped operators would induce an excess in $R_{K^{*}}$. In this work 
we only consider $C_{9}$ and $C_{10}$ for simplicity, although this can be extended to a general framework by combining $C_{9},C_{10} $  with their chirality flipped counterparts.
 
Assuming that the anomaly is generated by some new physics (up to possible standard model uncertainties), we will focus on $Z^{\prime}$ models and look at synergy between LHC constraints and $B$  physics using the language of simplified limits~\cite{Chivukula:2016hvp}.

\section{General $Z^{\prime}$ model}
Our objective is to use the data from LHCb along with high-energy results from CMS and ATLAS to identify the most compelling limits that can be set on $Z^{\prime}$ models capable of explaining the $R_{K}$ and $R_{K^{*}}$ anomalies.
We use the following model-independent parametrization\footnote{Here, following a simplified model analysis, for simplicity we ignore flavor-changing neutral-currents (FCNC) in the leptonic sector and do not consider neutrino couplings. As noted by the authors in \cite{Glashow:2014iga}, however, these constraints are likely to be important in any complete theory. 
 } for the coupling of an extra gauge boson $Z^{\prime}$ to fermions,
\begin{eqnarray}\label{lz}
\mathcal{L}_{Z^{\prime}}&=& \frac{1}{4}F^{\mu\nu}F_{\mu\nu}+ \frac{1}{2} m_{Z^{\prime}}^2 Z^{\prime \mu}Z^{\prime}_{\mu}  \nonumber \\
&+& \beta_{Z^{\prime}}Z^{\prime}_{\mu}\sum_{f= u,c,t}\sum_{\bar{f}^{\prime} =u,c,t}  
(c_{L}^{f^{\prime}f}\bar{f}^{\prime}\gamma^{\mu} P_L f+ c_{R}^{f^{\prime}f}\bar{f}^{\prime}\gamma^{\mu} P_R f)
\nonumber \\
&+&  \beta_{Z^{\prime}}Z^{\prime}_{\mu}\sum_{f= d,s,b}\sum_{f^{\prime} =d,s,b}  (c_{L}^{f^{\prime}f}\bar{f}^{\prime}\gamma^{\mu} P_L f +  c_{R}^{f^{\prime}f}\bar{f}^{\prime}\gamma^{\mu} P_R f)
\nonumber \\
 &+& \beta_{Z^{\prime}}Z^{\prime}_{\mu}\sum_{f= leptons}(c_{V}^{f}\bar{f}\gamma^{\mu} f + c_{A}^{f}\bar{f}\gamma^{\mu} \gamma_{5} f) ~,
\end{eqnarray}
where  $f, \bar{f}$ are SM fermions. The above parametrization 
therefore contains  chiral couplings to quarks ($c_{L},~c_{R}$), and vector and axial vector couplings ($c_{V},~c_{A}$) to leptons. 
Motivated by the LHCb results, we allow for flavor-changing neutral-currents in the quark sector and ignore them in the leptonic sector.
Finally, we use the normalization
\begin{equation}
\label{eq:def:betaz}
\beta_{Z^{\prime}} =m_{Z^{\prime}} \sqrt{\frac{\alpha_{em}V_{tb}V_{ts}^{\star} }{2  \pi v^2}} ~,
\end{equation}
so that (ignoring renormalization effects between the scales $M_{\Zp}$ and $m_b$)  the Wilson coefficients of the effective Lagrangian are related to the parameters of the $Z^{\prime}$ Lagrangian as follows,
\begin{eqnarray}\label{eq:c-def}
C_{9}^{\mu} \simeq c_L^{sb} c_V^{\mu}&,& \quad
C_{10}^{\mu} \simeq c_L^{sb} c_A^{\mu},  \\
C_{9}^{\prime \mu} \simeq c_R^{sb} c_V^{\mu}&,& \quad
C_{10}^{\prime \mu} \simeq c_R^{sb} c_A^{\mu} ~.
\end{eqnarray}
In the above, we have taken $V_{tb}~=~1$ and $V_{ts}=0.040$. 
One can invert these four equations and solve for $c_{L,R,V,A}$. Note, however, that in order to obtain a unique solution, one needs at least three of the Wilson coefficients to be non-zero.
Furthermore, in order to explain the anomaly with a single $Z^{\prime}$, we see that 

\begin{equation}\label{eq:c-ratio}
\frac{C^{\mu}_9}{C^{\prime\mu}_{9}} = \frac{C^{\mu}_{10}}{C^{\prime\mu}_{10}} ~,
\end{equation}

The analysis above sets constraints on predictions of any underlying model with a $Z^{\prime}$.  While the $Z^{\prime}$ must, at a minimum, 
couple to the bottom and the strange quark and decay to leptons to explain the anomaly, in a realistic model the $Z^{\prime}$ can couple to the 
light quarks (and possibly the top) as well. However, note that Eq. \ref{eq:c-ratio} implies that the value of $c^{sb}_{L}$ 
is not uniquely fixed by  the best fit values of $C_{9}$ and $C_{10}$. As we will show below, the value of $c^{sb}_L$ determines the overall strength of the $\Zp$ signal at the LHC in this minimal phenomenological model.
%We will choose $c_{L}$ to have a maximal value. 

In the next sections, we will explore the constraints on our $Z^{\prime}$ model arising from the LHCb observations, several theoretical considerations, and the dijet and dilepton searches for new resonances that are being conducted by the ATLAS and CMS collaborations.  We will use the language of simplified limits \cite{Chivukula:2016hvp} as a way to simultaneously express these different pieces of information and gain an overarching picture of the viable parameter space of the model.  First we will briefly review the key aspects of the simplified limits framework.  Then we will explore constraints on the simplest version of our $Z'$ model, in which its coupling to quarks comes only through a flavor-changing coupling to $sb$.  Following that, we see how the constraints are impacted if the $Z'$ also has a flavor-conserving coupling to quarks.

\section{Simplified limits in the Narrow Width Approximation}

Here, we will briefly review the simplified limits framework that was introduced in \cite{Chivukula:2016hvp} as a way to quickly understand how observing a new resonance in a particular production and decay channel could restrict the types of models  available to explain the observation.  As discussed below, this  framework will be useful for comparing several different kinds of information about our $Z'$ model, ranging from collider search data to the LHCb observations to  various theoretical bounds.

We will assume that the observed new resonance is narrow, such that any interference of standard model and new physics contributions can be neglected.
Thus the tree level partonic cross section for an $s$ channel narrow resonance $R$, decaying to a final state(s) 
$x+y$ from initial state partons $i +j$ can be written as, 
 \begin{equation}
\hat{\sigma}_{ij\to xy} = 16\pi (1 + \delta_{ij}) \frac{N_{S_{R}}}{N_{S_{i}}N_{S_{j}}} \frac{C_{R}}{C_{i}C_{j}}\Gamma (R ~\to~ x ~+~ y) \Gamma(R~\to i~+~j)\frac{\pi}{m_{R}\Gamma_{R}}\delta (\hat{s}-m_{R}^{2})~.
\end{equation}
In the above equation $N_{S}$ and $C$ count the number of spins and colors for the resonance $R$ and the incoming partons $i$ and $j$. 
The total cross section can be obtained by integrating the partonic cross section over parton
luminosities, and summing over incoming partons that contribute ({\it e.g.}, light quarks = $u,d,c,s$) and outgoing final states defining the signature of interest ({\it e.g.} light 
quarks for dijets, or $e$ and $\mu$ for dileptons),
\begin{equation}
\sigma = 16\pi^{2}\frac{N_{S_{R}}}{N_{S_{i}}N_{S_{j}}} \frac{C_{R}}{C_{i}C_{j}} \frac{\Gamma_{R}}{m_{R}}\times \Bigg(\sum_{ij} (1 + \delta_{ij})BR (R\to i ~+~j)\Bigg[\frac{1}{s}\frac{dL_{ij}}{d\tau}\Bigg]_{\tau=\frac{m_{R}^{2}}{s}}\Bigg)\Bigg( \sum_{xy} \Gamma(R\to x+y) \Bigg)~,
\label{eq:sigma-defn}
\end{equation} 
where the luminosity function $ \frac{dL_{ij}}{d\tau}$ is given by,\footnote{In this paper, for the purposes of illustration, we calculate these parton luminosities using the {\tt CT14NLO}~\cite{Pumplin:2002vw} parton density functions, setting the factorization scale $\mu_F^2= m^2_R$.}
\begin{equation}
\frac{dL_{ij}}{d\tau} = \frac{1}{1 + \delta_{ij} } \int_{\tau}^{1} \frac{dx}{x}\Bigg[f_{i}(x,\mu_{F}^{2})f_{j}(\frac{\tau}{x},\mu_{F}^{2}) + f_{j}(x,\mu_{F}^{2})f_{i}(\frac{\tau}{x},\mu_{F}^{2})\Bigg]~.
\end{equation}

It is useful to define the dimensionless ``simplified limits variable", $\zeta$, 
\begin{equation}\label{eq:def-zeta}
\zeta \equiv \frac{\Gamma}{m_{Z^{\prime}}}\sum_{production}(1+\delta_{ij}) BR (Z^{\prime} \to i + j) \sum_{decay}BR ( Z^{\prime} \to x + y)\ ~,
\end{equation}
which, as described more fully in \cite{Chivukula:2016hvp}, is a convenient general variable for expressing search
limits for narrow resonances. This variable is defined with respect to each production and decay channel and is  related to the effective size of the resonance-signal in that channel. 

In prior work, we showed how to use $\zeta$ as a tool for efficiently relating an observed signal of new physics to the predictions of entire classes of models at once. Here, we will use $\zeta$ as a common variable via which disparate constraints on a model may be compared.  We will both identify the range of $\zeta$ values our $Z'$ model must exhibit in order to explain the LHCb observations and analyze whether new particle searches at the ATLAS and CMS experiments are sensitive to that range of $\zeta$ values.  This will enable us to see whether our $Z'$ model remains viable in light of recent collider searches.

The form of the simplified limits variable for our $Z'$ boson in the dimuon decay channel may be obtained\footnote{As explained in \cite{Chivukula:2016hvp}, to interpret this ratio correctly one must also account for the different possible incoming partonic production mechanisms.}
 by applying Eqs. \ref{lz} and \ref{eq:def:betaz} to Eq. \ref{eq:def-zeta}:
 \begin{equation}\label{zeta}
\zeta^{\mu} = \left( \sum_{f,f^{\prime} = quarks} \frac{\beta^2_{\Zp}(|c^{f^{\prime}f}_{L}|^2 + |c^{f^{\prime}f}_{R}|^2 ) }{4\pi}\right)
\left(\frac{(|c^{\mu}_{A}|^2 + |c^{\mu}_{V}|^2) }{3\sum_{quarks}|c_{i}^{ff^{\prime}}|^2 + \sum_{leptons}|c_{i}^{j}|^2}
\right)~.
\end{equation}
In the denominator of the second factor, the subscript $i$ runs over left and right handed contributions for quarks, and over vector and axial vector contributions from leptons. 
A similar expression, $\zeta^j$ for the dijet signature, is easily found by replacing the numerator in the last factor by a sum over the relevant light-quark couplings. Thus $\zeta^{\mu}$ is quadratic in the undetermined parameter $c_{L}$, while $\zeta^j$ is quartic in $c_{L}$. 

In the next two sections we use this formalism to assess the $R_K$ anomaly in light 
of theoretical considerations and also  the CMS and ATLAS data gathered at 8 and 13 TeV.  \footnote{An analysis constraining 36 four fermion 
operators using dilepton data in the high $p_T$ tail was performed in \cite{Greljo:2017vvb}.}

\section{ Constraints on a $Z'$ coupling only to $sb$ and $\mu\mu$}\label{limiting-case}

We now consider how a $\Zp$-model that accomodates the $R_K$ anomalies would be constrained by theoretical information and by searches for dilepton and dijet resonances at ATLAS and CMS. For the purpose of illustration, we will evaluate $\zeta^{\mu}$ using representative Wilson coefficient values of $(C^{\mu}_{9}=-0.76 ,C^{\mu}_{10}=0.54,C^{\prime\mu}_{9}=0 ,C^{\prime\mu}_{10}=0)$ that are derived from fits performed using low energy data ~\cite{1704.05446}.\footnote{While different groups obtain larger significances depending on the fit parameters,  we chose the more conservative estimates of these Wilson coefficients \cite{1704.05446}.}  	

	Let us first consider a phenomenological model where the $\Zp$ couples to quarks only through the off-diagonal $sb$ coupling and also couples to muons but not other leptons.  This limiting case satisfies the minimum requirements in order for the $Z'$ to explain the $R_K$ anomalies and, as we will see, yields a wide range of allowable parameter space. We will  consider the more realistic case of a $\Zp$ with flavor-diagonal couplings to quarks, as well as the necessary $sb$ coupling, in  Section \ref{general-case}.\footnote{Having only off-diagonal $sb$ couplings, especially left-handed ones, would require very specific choices for the gauge-couplings and for the rotations required to translate between the gauge-eigenstate and mass-eignestate bases for the light fermions.  As we will see, having additional flavor-diagonal couplings to light quarks will generally enhance the dilepton and diquark signatures for the $\Zp$, and will correspondingly reduce the parameter space by current experimental constraints.}

	\begin{figure}
		\includegraphics[width= 0.5 \textwidth]{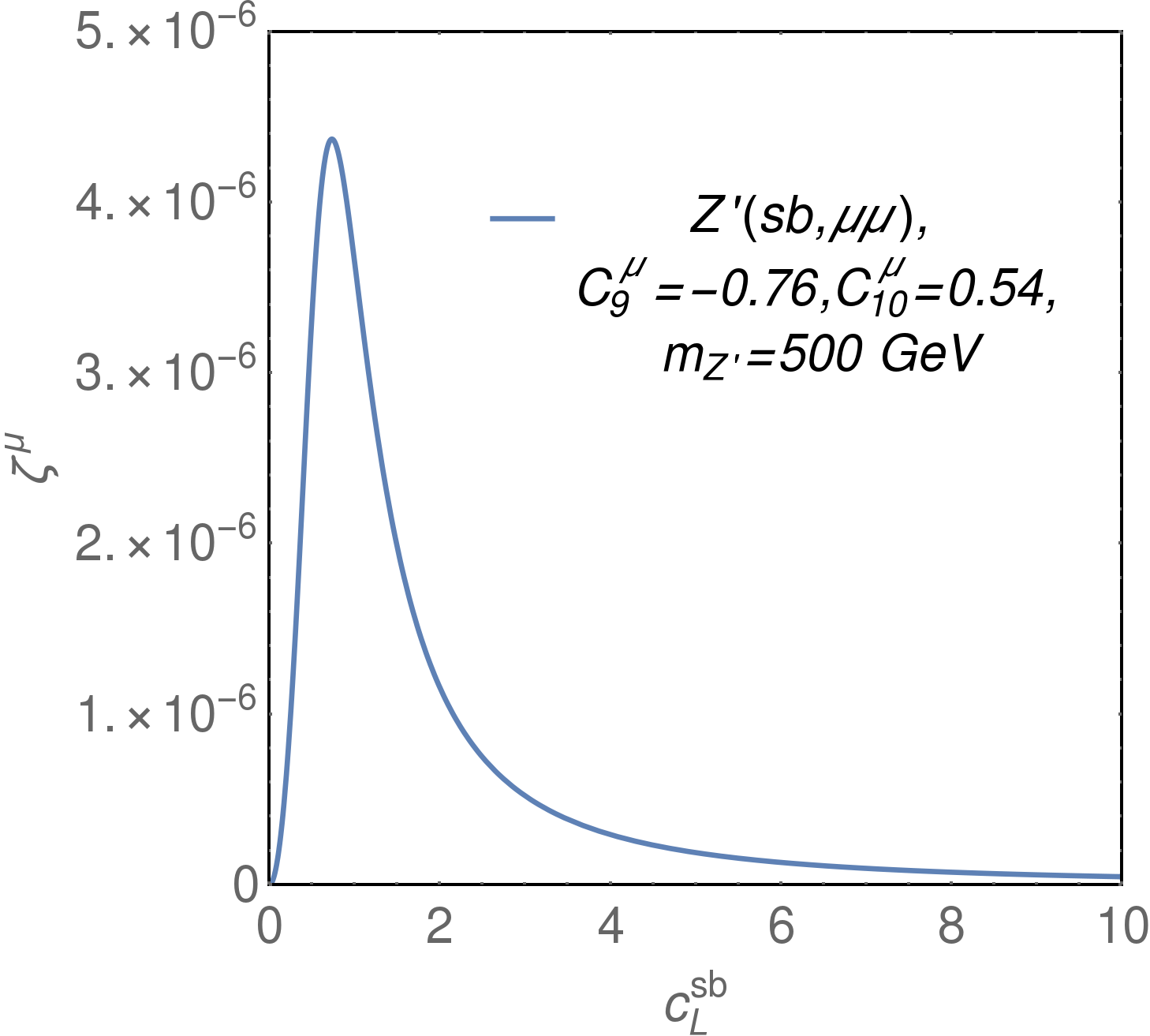}
		\caption{ The simplified limits variable $\zeta^\mu$ as a function of $c^{sb}_L$, for a $\Zp$ coupling only to $sb$ and muons; shown for the best fit values of $C_9,C_{10}$. The mass of the $\Zp$ only affects the normalization of this curve. The upper and lower bounds on $\vert c^{sb}_L \vert$ arise from requiring the $Z'$ couplings to remain perturbative. Note that the lowest value of $\zeta^\mu$ occurs for the smallest allowed value of $c^{sb}_L$.			\label{fig:cl-zeta}
		}
	\end{figure}

 \subsection{Upper limit on $\zeta^\mu$}
 
In the limiting case under consideration, the $\Zp$ is produced at the LHC only through $\bar{b}s + \bar{s}b$ luminosities. Henceforth, we will use the shorthand notation $qq^{\prime}$ to mean $(\bar{q}q^{\prime} + \bar{q}^{\prime}q)$. Under these assumptions, the form of $\zeta^\mu$ shown in Eq.~\ref{zeta} reduces to
{
\begin{eqnarray}\label{eq:zeta-simplest}
\zeta^\mu  &=&\beta^2_{\Zp} \frac{(c^{sb}_{L})^2  }{4\pi}
\left(\frac{(|c^{\mu}_{A}|^2 + |c^{\mu}_{V}|^2) }{3|c_{L}^{sb}|^2 + (|c^{\mu}_{A}|^2 + |c^{\mu}_{V}|^2)}
\right) \nonumber\\
&=&
\beta^2_{\Zp} \frac{(c^{sb}_{L})^2  }{4\pi}
\left(\frac{(C^{\mu}_{9})^2 + (C^{\mu}_{10})^2 }{3|c_{L}^{sb}|^4 + ( (C^{\mu}_{9})^2 + (C^{\mu}_{10})^2)}
\right)~.
\end{eqnarray}
}
 We present the relationship between $\zeta^\mu$ and $c^{sb}_{L}$ in Fig.~\ref{fig:cl-zeta} for $m_{\Zp}=500$ GeV, and setting $C_9, C_{10}$ at their best fit values. The only effect the mass has on the curve is to change the normalization of the $y-$axis. 
 
The curve in Fig.~\ref{fig:cl-zeta} shows that $\zeta^\mu$ has a maximum with respect to $c_{L}^{sb}$;  extremizing the expression for $\zeta^\mu$ in Eq. \ref{eq:zeta-simplest} identifies the value of $c_{L}^{sb}$ where the maximum occurs as
{
\begin{align}\label{eq:cL-max}
c_{L}^{sb} = \left\{\frac{1}{3} \big((C^{\mu}_{9})^2 + (C^{\mu}_{10})^2\big)\right\}^{1/4} \simeq 0.73\ .
\end{align}
}
Using this value in Eq. \ref{eq:zeta-simplest} reveals an upper bound on $\zeta$ given by,
{
\begin{align}\label{eq:zeta-max}
\zeta^\mu_{max} = \beta^2_{\Zp} \frac{\left\{\frac{1}{3} \big((C^{\mu}_{9})^2 + (C^{\mu}_{10})^2\big)\right\}^{1/2}}{8 \pi}\ .
\end{align}
}
The value of $\zeta^\mu_{max}$ determines the maximum possible size of the dimuon signal that our $Z'$ could produce at ATLAS and CMS.  To see this, compare Eqs. \ref{eq:sigma-defn} and \ref{eq:def-zeta}; one can clearly write $\zeta^\mu$ in terms of the size of the peak cross-section for producing the resonance.  As described in detail in \cite{Chivukula:2016hvp}, the precise relationship is:
\begin{align}
\zeta =\frac{\sigma} { 16\pi^2 \cdot \frac{N_{S_{R}}}{N_{S_{i}}N_{S_{j}}} \times 
\left[\sum_{ij} \omega_{ij} \left[\frac{1}{s} \frac{d L^{ij}}{d\tau}\right]_{\tau = \frac{m^2_R}{s}}\right]} ~.
\label{eq:zeta-sigma}
\end{align}
where the weighting factor $\omega$ is defined as
\begin{equation}
\omega_{ij} \equiv \dfrac {(1 + \delta_{ij})  BR(R\to i+j)} {\sum_{i'j'} (1 + \delta_{i'j'}) BR(R\to i'+j')}~.
\end{equation}
Since the parton luminosities and the spin and color factors are fixed, the maximum $\zeta^\mu$ does indeed correspond to a maximum value of $\sigma$.

\subsection{Lower limit on $\zeta^\mu$ from Perturbativity}

Retaining our present focus on a $Z'$ with only off-diagonal coupling  to  quarks, the minimum size of $\zeta^\mu$ that is consistent with the LHCb observations is determined by several  theoretical considerations. The first constraint we assess is that of perturbativity of this phenomenological $\Zp$ model. 

	Perturbativity requires that all quark and lepton couplings be bounded from above. Based on the notation in Eq. \ref{lz}, we will use the estimates:
	\begin{align}
	\beta_\Zp\times \vert c_{L}^{sb}\vert < 4 \pi \ \ \ \ {\rm and} \ \ \ \ \beta_\Zp\times \vert c_{V}^{\mu}\vert < 4 \pi
	\end{align}
	to establish the conditions under which perbativity exists.  The first of these tells us directly that $\vert c_{L}^{sb}\vert < 4 \pi / \beta_{\Zp}$.  To interpret the second, we note that from Eq. \ref{eq:c-def} we can write,
	\begin{align}
	c_V^\mu= \frac{C_9}{c_{L}^{sb}},\quad c_A^{\mu} = \frac{C_{10}}{c_{L}^{sb}}\ .
	\end{align}
	 Therefore, from the benchmark value of $C_{9}$, we can derive a lower limit\footnote{For our benchmark values, the smaller magnitude of $C_{10}$ would yield a weaker constraint.} on the value of $\vert c_L^{sb} \vert$ to pair with the upper limit mentioned just above:
	\begin{align}\label{eq:zeta-low-perturbative}
	\frac{0.76\beta_{\Zp} }{4 \pi} < \rm |c_{L}^{sb}| < \frac{4 \pi }{\beta_{\Zp}}
	\end{align}
Incorporating information from Eq. \ref{eq:def:betaz}, we find%
\begin{align}
 1.73\times 10 ^{-6}~ [\rm GeV]^{-1}~ m_{\Zp}\ \  < \ \ \ |c_{L}^{sb}|\ \ \  < \ \ \frac{4.4\times 10^{5}~[\rm GeV]}{m_{\Zp}} \,.
	\end{align}
Note that the allowed range of values of $\vert c_L^{sb} \vert$ reduces in scope with increasing $m_{\Zp}$. 

As may be seen from the shape of the curve plotted in Fig.~\ref{fig:cl-zeta}, for a given mass $m_{\Zp}$ the lower bound on $c^{sb}_L$ that we have just derived yields a {\it lower bound} on $\zeta^\mu$, which we will denote $\zeta^\mu_{pert}$.

{
	Finally, we note that the flavor changing neutral currents mediated by the $\Zp$ affect the $B^{0}_{s} - \bar{B}^{0}_{s}$ system and contribute to mixing of the states through a tree level diagram.
	It is possible to use the measured mass difference  in the $B^{0}_{s} - \bar{B}^{0}_{s}$ system to determine an upper bound on $c^{sb}_L$ \footnote{For a description of the details of this calculation, see for example ref~\cite{Alok:2017jgr}}. We find that $|c_L^{sb}| \lesssim 4.4 \times 10^4$ at 95\% confidence level. For resonances with masses in the TeV range, as considered here, this limit is much weaker than the upper bound derived above from perturbativity arguments.	}

%%%%%%%%%%%%%%%%%%%%%%%%%%%%%
	\subsection{Lower limit on $\zeta^\mu$ Due to the $Z'$ Width}
	For a $Z'$ boson coupling only to the $s$, $b$ and $\mu$, the expression for the decay width is,
	{ 
	\begin{align}\label{eq:width}
	\Gamma= \beta_{\Zp}^2 \left[
	\frac{1}{12 \pi}\left(
	\frac{(C_{10}^{\mu})^2}{(c_L^{sb})^2} + \frac{(C_{9}^{\mu})^2}{(c_L^{sb})^2}
	\right)
	+
	\frac{(c_{L}^{sb})^2}{4 \pi}
	\right] m_{\Zp}
	\end{align}
}
	\begin{figure}
		\includegraphics[width =0.5 \textwidth]{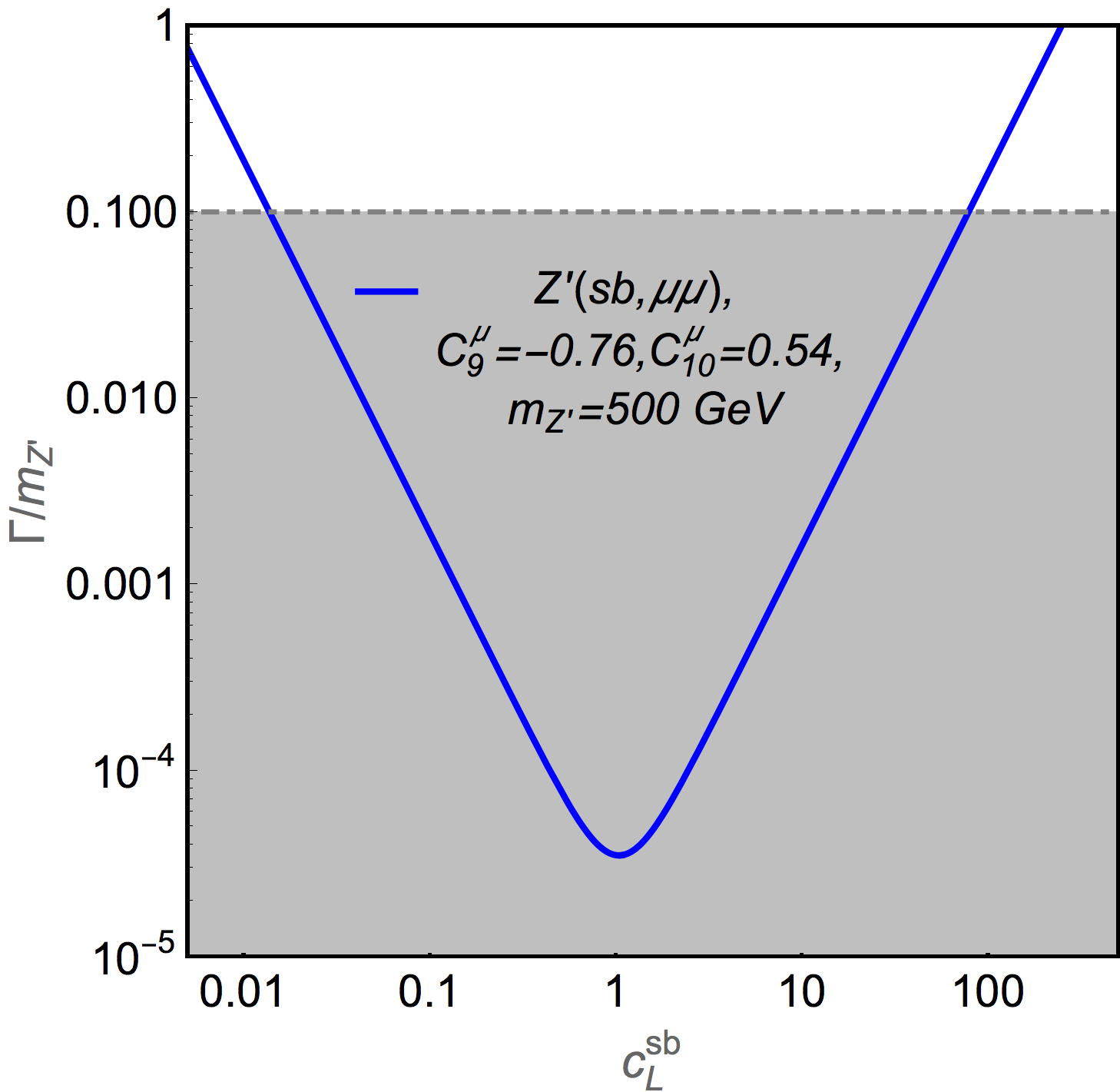}
		\caption{Dependence of $\Gamma/m_{\Zp}$ upon $c_{L}^{sb}$. The shaded region corresponds to $Z'$ boson with a narrow width, $\Gamma/M < 0.1$.}\label{fig:cl-R}
		\end{figure}
	As noted earlier, we will focus on narrow resonances, defined as those satisfying $\Gamma/M \leq \alpha_{max}$ for a value of $\alpha_{max}$ that we will specify below. Restricting our attention to narrow resonances provides constraints on the value of $c_{L}^{sb}$ given by,
	\begin{align}
	\frac{4\pi \alpha_{max}}{2 \beta_{\Zp}^2}\left(
	1 - \sqrt{D}
	\right)<(c_{L}^{sb})^2 < \frac{4\pi \alpha_{max}}{2 \beta_{\Zp}^2}\left(
	1 + \sqrt{D}
	\right)
	\end{align}
 where  the discriminant D is defined as 
{ 
 \begin{align}
   D\equiv \left[1 - \frac{\beta_{\Zp}^4 }{12 \pi^2 (\alpha_{max})^2}. \left((C_{10}^{\mu})^2 + (C_{9}^{\mu})^2\right)  \right]
   \end{align}
}
with $\Gamma/ m_{\Zp} = \alpha_{max}$. Values of $c^{sb}_L$ falling below the lower bound, though potentially consistent with the perturbativity constraint in Eq.~\ref{eq:zeta-low-perturbative}, would nonetheless yield $\Gamma/M > \alpha_{max}$. So this provides a stricter lower bound on $c^{sb}_L$.  We will denote the value of $\zeta^\mu$ corresponding to this lowest value of $c_L^{sb}$ by the symbol $\zeta^\mu_{narrow}.$
		
		Fig.~\ref{fig:cl-R} displays the dependence of $\Gamma/m_{\Zp}$ on $c_L^{sb}$. The shaded region corresponds to the values of $c_{L}^{sb}$ where the $\Zp$ has a narrow width, which we hearafter specifically take to mean $\Gamma/M < \alpha_{max} = 0.1$.  In addition to the minimum value of $c_L^{sb}$ for which the $Z'$ boson is still ``narrow", we can also identify in the figure a unique value  of $c_L^{sb}$ at which $\Gamma/M$ is minimized. Comparing Eqs. \ref{eq:zeta-simplest} and \ref{eq:width} shows that for fixed $C_{9,10}$ the width is inversely proportional to $\zeta^\mu$, so that the value of $c^{sb}_L$ at which $\Gamma/M$ is minimized also corresponds to $\zeta^\mu_{max}$ in Eq. \ref{eq:cL-max}.
 
Furthermore, since the resonance width in Eq.~\ref{eq:width} is proportional to the cube of the mass, the solid curve in Fig.~\ref{fig:cl-R} moves upward as the mass of the $Z'$ increases (assuming fixed values of $C_9^\mu$ and $C_{10}^\mu$).  For a heavy enough $Z'$, only the lowest point of the solid curve will still be within the shaded region; i.e., only for that unique value of $c_L^{sb}$ will the resonance still be narrow.  This happens when the discriminant $D$ becomes negative, which allows us to find the corresponding maximum value of the $Z'$ mass (shown for $\alpha_{max}=0.1$).
{
	\begin{align}
	m_{\Zp}< \sqrt{\frac{4 \sqrt{3}\pi^2 v^2 \alpha_{max}}{\left((C_{10}^{\mu})^2 + (C_{9}^{\mu})^2 \right)^{1/2}\alpha_{em} V_{tb}V_{ts}^{\star}}} \simeq 37.8~\text{TeV}\ .
	\end{align}
}

The interesting range of the simplified limits variable $\zeta^\mu$ for this minimal model where the $Z'$ boson couples only to $s$, $b$, and $\mu$ is therefore $\zeta^\mu_{pert,narrow} < \zeta^\mu<\zeta^\mu_{max}$. We will now consider what the ATLAS and CMS resonance searches in the dilepton and dijet channels can say about that region of parameter space.

	%%%%%%%%%%%%%%%%%%%%%%%%%%%%%%%%%%%%%%%%%%
	\subsection{Upper limit on $\zeta^\mu$ from dilepton resonance searches.}

The ATLAS and CMS experiments are searching for new resonances decaying to dileptons, and therefore can potentially constrain or discover the $Z'$ boson proposed here.  As discussed in \cite{Chivukula:2016hvp}, the simplified limits variable $\zeta^\mu$ provides a useful way to report the results of  such searches.  Here, we will find that it also enables us to overlay several different kinds of information about our $Z'$ state.

\begin{figure}[t]
	\centering
\includegraphics[width= 0.6 \textwidth]{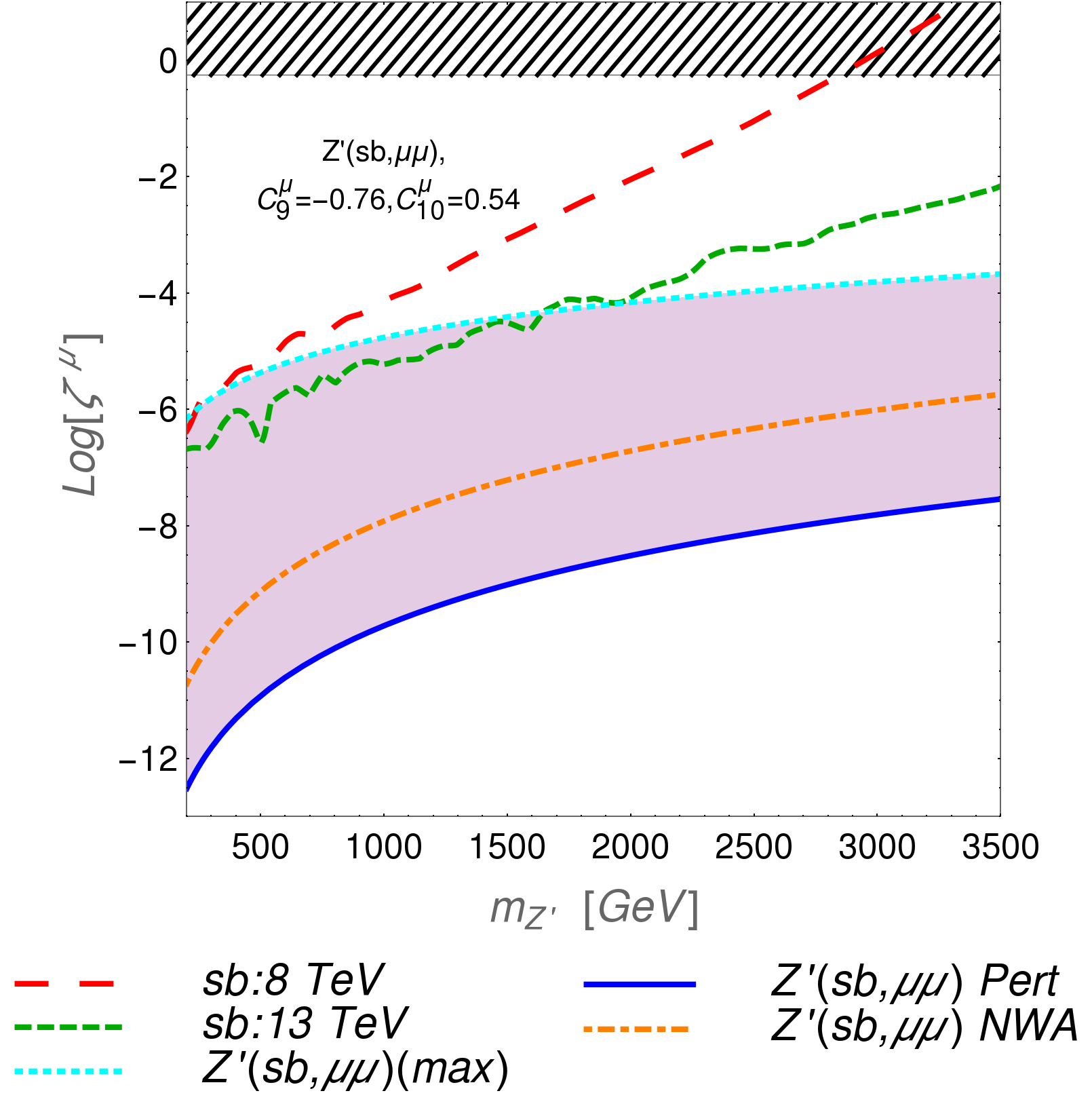}
\caption{Constraints on the $Z^{\prime}$ coupling only to $s$, $b$, and $\mu$.  The  pink shaded region shows where $\zeta^\mu$ is small enough (below light-blue dashed line) to  explain the $R_K$ and $R_{K^*}$ anomalies and also large enough (above blue solid line) to keep the $Z'$ boson's couplings perturbative.  Above the orange dot-dashed line, the $Z'$ would be narrow.  The maximum upper bounds (from the absence of observing dilepton resonances) on $\zeta^\mu$ derived from  both the 8 TeV (red long-dashed) and 13 TeV (green dashed) ATLAS dilepton limits~\cite{Aad:2014cka,Aad:2014wca,ATLAS-CONF-2017-027} are shown for comparison. We see that LHC dilepton results are only beginning to be sensitive to the region which can explain the $R_K$ anomalies. }
\label{fig:limits}
\end{figure}

We use both 8 TeV and 13 TeV dilepton resonance searches at the LHC to extract limits on the model~\cite{Aad:2014wca,Aad:2014cka}.  Specifically, we use Eq. \ref{eq:zeta-sigma} to reframe the experimental limits on $\sigma\cdot B \cdot {\cal A}$ as upper bounds on $\zeta^\mu$, following the methods of \cite{Chivukula:2016hvp}. Fig. \ref{fig:limits} displays the resulting constraints in the $\log \zeta^{\mu}$--$M_{\Zp}$ plane. The red long-dashed line corresponds to the upper limit on the value of $\zeta^{\mu}$ from dilepton constraints assuming $sb$-initiated production at 8 TeV, while the green dashed line shows how that bound is strengthened by the data taken at 13 TeV.

The pink shaded curved band represents the region of $\zeta^\mu$ values that simultaneously explain the 
$R_{K}$ anomalies (i.e., fall below the light-blue dotted line) and remain consistent with perturbativity (i.e., lie above the solid blue curve). Points lying above the orange dot-dashed line within the shaded band correspond to cases where the $Z'$ resonance is narrow.

At present, the LHC dilepton resonance searches leave this allowed region of $\zeta^\mu - m_\Zp$ space essentially intact for $Z'$ boson that couples only to $s$, $b$, and $\mu$. The 8 TeV data provides an upper bound on $\zeta^\mu$ for $Z'$ masses below {300} GeV.  The 13 TeV data, however, is able to probe further into the upper edge of the pink band for a $Z'$ mass below {1.7} TeV, and excludes some of those values of $\zeta^\mu$. We anticipate that future LHC dilepton data will explore the region defined by the pink band more thoroughly, thereby testing our $Z'$ model's viability as an explanation of the $R_K$ anomalies.

In principle,  one can also use dilepton data to extract the flavor non-universality limits by comparing the cross-sections in the di-electron channel versus the di-muon channel~\cite{1606.01736} in the high mass Drell-Yan data. Using the narrow width approximation, as described above, the interference of the $Z'$ with the Drell-Yan background can be neglected. Therefore,  the predicted flavor cross-section ratio is given by:
\begin{align}
	\frac{\sigma\left(pp\rightarrow \mu^+\mu^-\right)}{\sigma\left(pp\rightarrow e^+e^-\right)} = 1 + \frac{\sigma_{Z'}}{\sigma_{\text{SM}}}.
\end{align}
	Due to a lack of information about the uncertainities of this measurement, however, 
	we do not use the existing data to impose a limit on the $\Zp$ model via this method. We expect this will become a valuable line of inquiry as future data emerges.

	%%%%%%%%%%%%%%%%%%%%%%%%%%%%%%%%%%%%%%%%%%
	\subsection{Upper limit on $\zeta^j$ from dijet resonance searches.}
	\label{sec:dijet}

Searches  for new physics in dijet final states have been conducted by the ATLAS and CMS collaborations \cite{CMS-PAS-EXO-16-056,Aaboud:2017yvp,Aad:2014aqa}. In this section we will show the relation between the constraints derived from dijet resonance searches and those from the dilepton resonance searches discussed earlier. 

Following the reasoning used to define $\zeta^\mu$ for the process $sb \to \Zp \to \mu\mu$, we can analogously define the variable $\zeta^j$ for $sb\to \Zp \to jj$:
{
\begin{eqnarray}\label{eq:zeta-simplestj}
\zeta^j  &=&\beta^2_{\Zp} \frac{3(c^{sb}_{L})^4  }{4\pi}
\left(\frac{1 }{3|c_{L}^{sb}|^2 + (|c^{\mu}_{A}|^2 + |c^{\mu}_{V}|^2)}
\right) \nonumber\\
&=& \beta^2_{\Zp} \frac{3(c^{sb}_{L})^6  }{4\pi} \left( \frac{1}{3|c_{L}^{sb}|^4 + (|C^{\mu}_{9}|^2 +  |C^{\mu}_{10}|^2)} \right)
\end{eqnarray}
}
Note that $\zeta^{j}$ is a monotonically increasing function of $c_L^{sb}$; hence there is no equivalent for $\zeta^j$ of the lower bound we found for $\zeta^\mu$ based on maintaining perturbativity of the $Z'$ boson's couplings to fermions or narrowness of the resonance's width. Moreover, even though $\zeta^\mu$ is generally double-valued in $c_L^{sb}$, its maximum value, $\zeta^\mu_{max}$, corresponds to a unique value of $c_L^{sb}$ (as in Fig.~\ref{fig:cl-zeta} and Eq. \ref{eq:cL-max}); it therefore corresponds to a unique value of $\zeta^j$ through the formal relationsip {$\zeta^j = 3\zeta^\mu (c_L^{sb})^4 / (\vert C_9^\mu\vert^2 + C_{10}^\mu\vert^2)$} that is implied by Eqs. \ref {eq:zeta-simplest}
 and \ref{eq:zeta-simplestj}.  We will use this in relating the various collider limits to one another.

\begin{figure}[t]
	\centering
\includegraphics[width= 0.57 \textwidth]{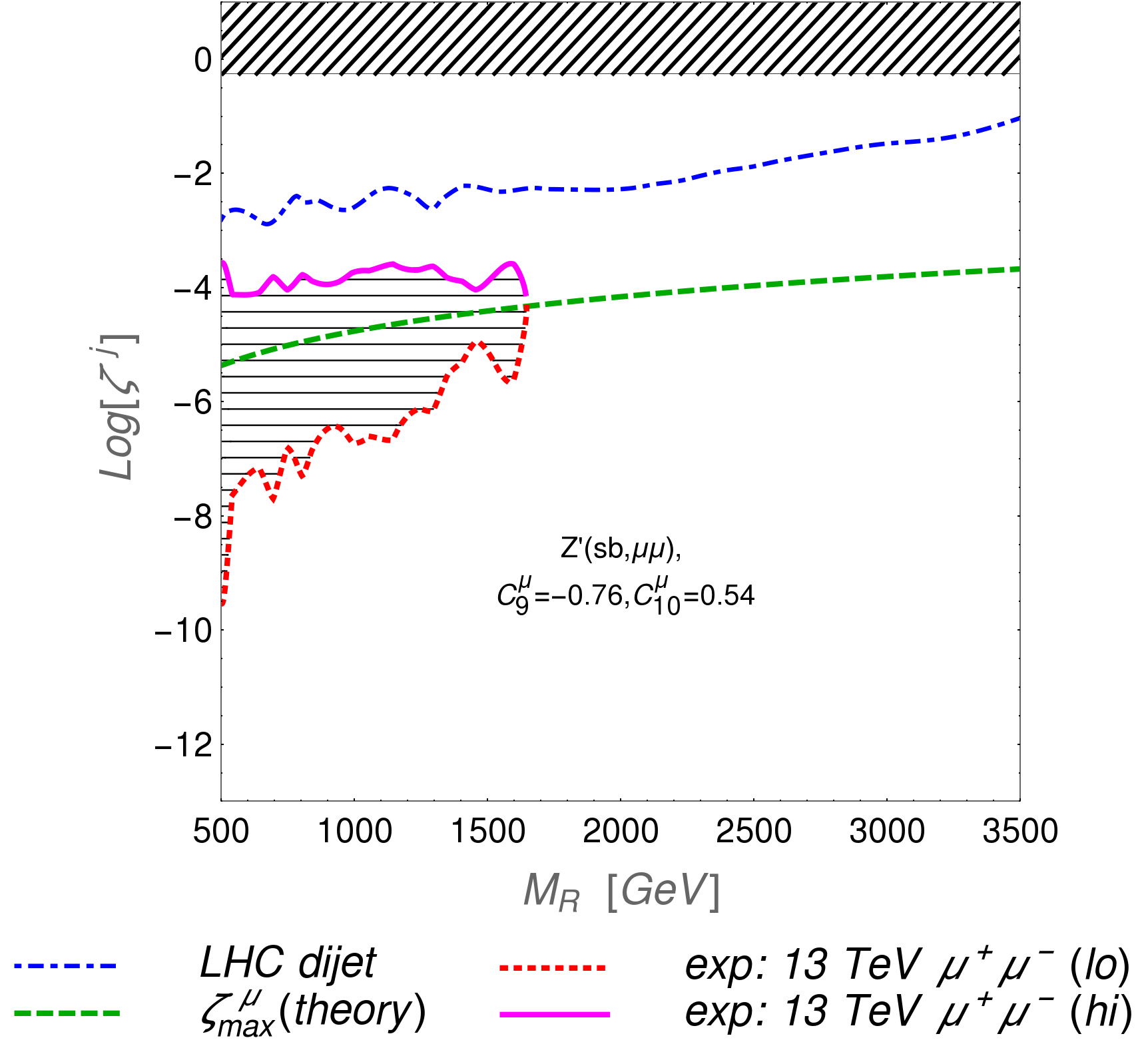}
\caption{Constraints on the $Z^{\prime}$ model in the $\log{\zeta^j}$ vs mass plane. The blue (dot-dashed) line is the maxium allowed upper bound on $\zeta^j$ from the absence of observed dijet reosonances by CMS at $13$~TeV energy~\cite{CMS-PAS-EXO-16-056}. The green (dashed) line indicates the value of $\zeta^j$ corresponding to $\zeta^\mu_{max}$, associated with a unique value of $c^{sb}_L$. Both {\it higher and lower} values of $\zeta^j$ are allowed by moving away from that specific value of $c^{sb}_L$. The horizontally hatched region on the left edge indicates values of $\zeta^j$ that are ruled out by dilepton searches as consistent with our benchmark values of $C^\mu_9 $ and $C^\mu_{10}$.  }\label{fig:limits-dijet}
\end{figure}

In Fig.~\ref{fig:limits-dijet} the blue (dot-dashed) line shows the upper bound on $\zeta^j$ imposed by  dijet resonance searches using $36\, \text{fb}^{-1}$ of CMS data at a center of mass energy of $13$~TeV~\cite{CMS-PAS-EXO-16-056}.~\footnote{Note that the dijet and dilepton bounds are exclusions at 
95 \% C.L (2 $\sigma$). }
The green (dashed) line shows the value of $\zeta^j$ corresponding to $\zeta^\mu_{max}$. The horizontally hatched area at the left edge shows the region of $\zeta^j$ and $M_R$ that are ruled out by the dilepton resonance searches discussed earlier. As shown in 
Fig.~\ref{fig:cl-zeta}, each value of $\zeta^\mu$ (except $\zeta^\mu_{max}$) corresponds to two different values of $c^{sb}_L$. Therefore, as the ATLAS data impose an upper bound on $\zeta^\mu$, some intermediate range of $c^{sb}_L$ values is eliminiated, leaving both smaller and larger values of $c^{sb}_L$ still allowed.  As $\zeta^j$ is a monotonically increasing function of $c^{sb}_L$, this results in eliminating a range of $\zeta^j$ values to either side of the $\zeta^\mu_{max}$ curve in Fig.~\ref{fig:limits-dijet}. The hatched region stops abruptly at about {1.7} TeV because, as shown in Fig.~\ref{fig:limits}, the dilepton experimental constraints on $\zeta^\mu$ only are only sensitive to the theoretically interesting range of $\zeta^\mu$ for $Z'$ bosons with masses below that value.

\section{Constraints on a $\Zp$ coupling to all fermions}\label{general-case}

So far we have considered the case when the $\Zp$ couples only to strange quarks, bottom quarks, and  muons. We now generalize the constraints discussed above to the situation when the $\Zp$ couples to all fermions.\footnote{Generically, we see from Eq. \ref{eq:def-zeta} that the addition of bosonic decay channels will reduce the value of $\zeta^j$, weakening somewhat the limits dicussed below. Qualitatively, however, the features described here remain unchanged.}

To make our discussion more tractible, we define a simplified notation as follows:
{
\begin{align}
&c_{q}^2 \equiv \frac{1}{2} \sum_{f=u,d,s,c,b} \left(|c_{L}^{ff}|^2 + |c_{R}^{ff}|^2 \right)+ \sum_{ff^{\prime}=\left\{uc,ds,db\right\}}\left(|c_{L}^{ff^\prime}|^2 + |c_{R}^{ff^\prime}|^2\right), &
\end{align}
\begin{align}
c_{t}^2 \equiv \frac{1}{2}\sum_{f=t} \left(|c_{L}^{ff}|^2 + |c_{R}^{ff}|^2 \right)+ \sum_{ff^{\prime}=\left\{ut,ct\right\}}\left(|c_{L}^{ff^\prime}|^2 + |c_{R}^{ff^\prime}|^2\right)\ . 
\end{align}
}
Thus $c_{q}$ corresponds to all possible partonic combinations in $pp$ collisions\footnote{We neglect the top quark pdfs inside the proton. We also ignore top mass effects in the decay width for the time being to simplify our discussion.}. 
We also define,
{
\begin{align}
&c_{e}^2 \equiv |c_{V}^{e}|^2 + |c_{A}^{e}|^2& \\
&c_{\tau}^2 \equiv |c_{V}^{\tau}|^2 + |c_{A}^{\tau}|^2& \\
&c_{\mu}^2 \equiv |c_{V}^{\mu}|^2 + |c_{A}^{\mu}|^2 = \frac{\left((C^\mu_9)^2 + (C^\mu_{10})^2\right)}{|c^{sb}_{L}|^2} + \frac{\left((C^{\prime \mu}_9)^2 + (C^{\prime\mu}_{10})^2\right)}{|c^{sb}_{R}|^2}& \label{eq:gen-cmu-def}
\end{align}
}
 For the ease of discussion we again set $C^{\prime \mu}_9 =C^{\prime \mu}_{10} =0 $, so that $c_{R}^{sb} = 0$. Additionally, we also set $c_{e}=0$, and choose non-zero values only for $c_\tau$ and $c_q$. 
 With the above assumptions $\zeta^{\mu}$ is given by,
 {
\begin{align}
	&\zeta^{\mu}=\left(\sum \frac{ \beta^2_{\Zp}(|c^{f^{\prime}f}_{L}|^2 + |c^{f^{\prime}f}_{R}|^2 ) }{4\pi}\right)
	\left(\frac{(|c^{\mu}_{A}|^2 + |c^{\mu}_{V}|^2) }{3\sum_{quarks}|c_{i}^{j}|^2 + \sum_{leptons}|c_{i}^{j}|^2}
	\right) \nonumber &\\
	& =\frac{\beta^2_{\Zp}}{4\pi} (c_{q}^2 + (c_{L}^{sb})^{2})\frac{(c_{\mu}^2 )}{3 c_{q}^2 + 3 (c^{sb}_{L})^2 +  c_{\mu}^2 + c_{\tau}^2  }.&\label{eq:gen-zeta-mu-def}
\end{align}
}

The maximum value of $\zeta^{\mu} $ now corresponds to the following value of $c_{L}^{sb}$:
{
\begin{align}
c_{L}^{sb}=\left(\frac{ \sqrt{(C^\mu_{10})^2+(C^\mu_{9})^2-c_\tau^2 c_q^2}}{\sqrt{3}}-c_q^2\right)^{1/2}.\label{eq:gen-cl-max-def}
\end{align}
}
Since $\zeta^\mu$ is a real quantity, we observe that its maximum allowed value corresponds to $c_{L}^{sb} = 0$ in two situations: either when $c_\tau^2 c_q^2 > (C^\mu_{10})^2 + (C^\mu_{9})^2$, or when {$\frac{ \sqrt{(C^\mu_{10})^2+(C^\mu_{9})^2-c_\tau^2 c_q^2}}{\sqrt{3}} < c_q^2$}. 

\subsection{Upper limit on $\zeta^\mu$}

By applying Eqs. \ref{eq:gen-cmu-def}  and \ref{eq:gen-cl-max-def}
 to Eq. \ref{eq:gen-zeta-mu-def}, we obtain a general expression for the maximum value of $\zeta^\mu$ 
 {
\begin{align}
\zeta_{max}^{\mu} = \frac{\beta_{\Zp}^2 \left((C^\mu_{10})^2+(C^\mu_{9})^2\right)}{4\pi  \left(2 \sqrt{3} \sqrt{\left((C^\mu_{10})^2+(C^\mu_{9})^2-c_{\tau}^2 c_{q}^2\right)}+4 c_{\tau}^2-3 c_{q}^2\right)}.
\end{align}
}
From this equation, we see that introducing only $c_\tau$ (and setting $c_q = 0$), reduces the maximum value of $\zeta$; i.e., adding decay modes that do not contribute to $Z'$ production reduces $\zeta^\mu_{max}$.  Conversely,if we set $c_\tau = 0$ then allowing $c_q \ne 0$ increases the maximum value of $\zeta^\mu$; adding decay modes that contribute to production (beyond the minimum required production via $sb$ annihilation) increases the  value of $\zeta^\mu_{max}$. Keep in mind, however, that if $((C^\mu_{10})^2+(C^\mu_{9})^2) < c_{\tau}^2 c_{q}^2$, then $c_{L}^{sb}=0$, and without the flavor-diagonal coupling to quarks, our $Z'$ boson would not be able to explain the $R_K$ anomalies. 

\subsection{Lower limit on $\zeta^\mu$ from Perturbativity}

While the perturbativity constraints on $c_{L}^{sb}$ itself remain unchanged, the expression for the minimum value of $\zeta^{\mu}$ that is allowed by perturbativity is altered.
The general form of $\zeta^{\mu}$  can now be written as, 
{
\begin{align}
\zeta^{\mu} =\frac{\beta_{\Zp}^2 \left(C_{10}^2+C_{9}^2\right) \left((c_{L}^{sb})^2+c_{q}^2\right)}{4\pi  \left( C_{10}^2+ C_{9}^2+(c_{L}^{sb})^2 \left( c_{\tau}^2+3 \left((c_{L}^{sb})^2+c_{q}^2\right)\right)\right)}.\label{eq:pertgeneral}
\end{align}
}
Comparing this with the form of Eq.~\ref{eq:zeta-simplest} shows that the constraint on perbutativity is weaker than in the situation where the $\Zp$
couples only to bottom and strange quarks.

\subsection{Lower limit on $\zeta^\mu$ Due to the $Z'$ Width}\label{sec:width}
For the generalized scenario the decay width is given by,\footnote{Expressions for the partial widths are given in Appendix \ref{ap1}.}
{\begin{align}
\Gamma= \beta_{\Zp}^2 \left[
\frac{1}{12 \pi}\left(
\frac{(C_{10}^{\mu})^2}{(c_L^{sb})^2} + \frac{(C_{9}^{\mu})^2}{(c_L^{sb})^2 }+ c_\tau^2
\right)
+
\frac{(c_{L}^{sb})^2 + c_ q^2  + c_t^2}{4 \pi}
\right] m_{\Zp}
\end{align}
}
Comparing this with Eq.~\ref{eq:width}, indicates that the resulting constraint on $\zeta^\mu$ will be stronger than was the case for the $Z'$ boson coupling only to $s$, $b$ and $\mu$.

\subsection{Upper limit on $\zeta^\mu$ from dilepton resonance searches.}

\begin{figure}[t]
	\centering
\includegraphics[width= 0.57 \textwidth]{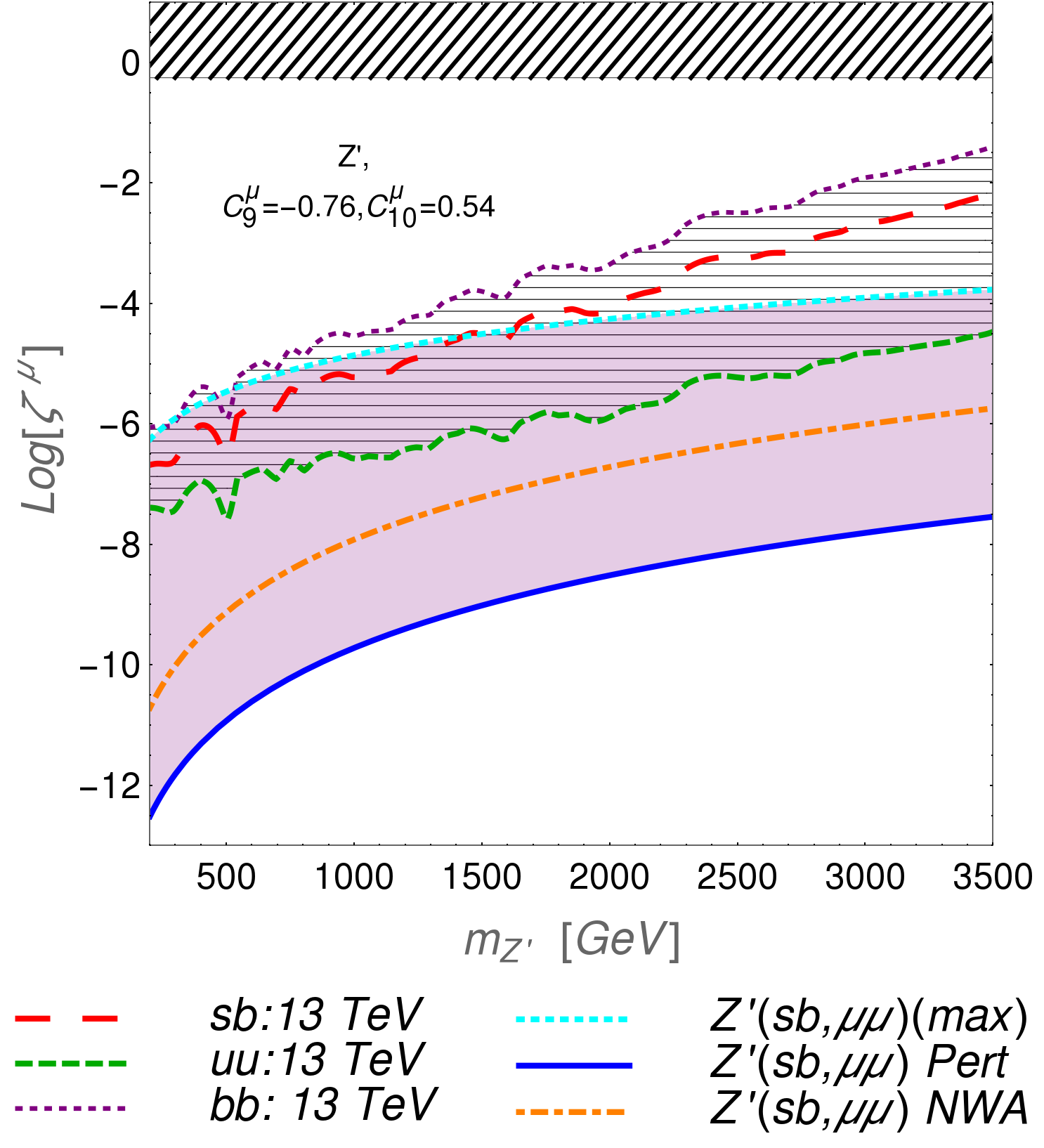}
\caption{Constraints on the general $Z^{\prime}$ boson that couples to all fermions.  The  pink shaded region reproduced from Fig. 3 (see text for discussion of how this region shifts for a general $Z^{\prime}$) shows where $\zeta^\mu$ is small enough (below light-blue dashed line) to  explain the $R_K$ and $R_{K^*}$ anomalies and also large enough (above blue solid line) to keep the $Z'$ boson's couplings perturbative.  Above the orange line, the $Z'$ would be narrow. The upper bounds on $\zeta^\mu$ derived from the 13 TeV ATLAS dilepton limits~\cite{Aad:2014cka,Aad:2014wca,ATLAS-CONF-2017-027} for production via $sb$ (red large-dashed), $uu$ (green dashed), and $bb$ (purple dotted) initial states are shown for comparison. }
\label{fig:limits-general}
\end{figure}

The ATLAS dilepton resonance searches at 13 TeV~\cite{Aad:2014cka,Aad:2014wca,ATLAS-CONF-2017-027} give rise to upper bounds on $\zeta^\mu$ as described earlier; our results for the $Z'$ coupling to all fermions appear in Fig. \ref{fig:limits-general}. The location and shape of each limit curve in the $\log[\zeta^\mu] - m_{Z'}$ plane depend on one's assumptions about the dominant production mechanism for the $Z'$ state.  We capture the range of posssibilities by showing a curve that assumes the $Z'$ is primarily produced by $b\bar{b}$ annihilation (which gives a relatively weak constraint due to the smaller parton luminosity involved) and another assuming that production via $u\bar{u}$ dominates.  The horizontally-hatched region between these curves is the region within which the upper bound on $\zeta^\mu$ for any $Z'$ boson coupling to a combination of $b\bar{b}$ and $u\bar{u}$ will fall.  For comparison, we also show the curve assuming that the $Z'$ is produced only through an off-diagonal $sb$ coupling.

  Fig. \ref{fig:limits-general} also shows, as a pink band, the swath of parameter space of greatest interest for our simplest model.  The upper border (light-blue dotted line) deliniates the area that can explain the $R_K$ anomalies;  there is an absolute lower limit originating from perturbativity (blue-solid line), and a  comparatively softer constraint if we demand that the resonance is narrow (orange dashed) line. For this $Z'$ boson coupling to all fermion flavors, the dilepton search limits are potentially able to rule out a significant fraction of the parameter space that can explain the $R_K$ anomalies. 
  
 Note, however, that the precise location and size of the pink band in the general model will vary depending on which additional $Z^\prime$ production and decay modes become available.  Relative to the top of the pink band from the minimal case that is reproduced in Fig. \ref{fig:limits-general}, the upper edge denoting the upper limit on $\zeta^\mu$ moves down (up) when additional production (decay) modes are included in the model.  The location of the bottom edge of the pink band (from perturbativity constraints) moves down when any additional modes are added, thereby increasing the size of the pink band.  The location of the orange dashed line (denoting a narrow resonance) always moves upwards relative to what is shown in Fig.~\ref{fig:limits-general}, giving rise to stronger constraints; details depend on how many additional non-zero couplings are present.

Searches for new resonances decaying to dijets yield constraints on $\zeta^j$ similar to those illustrated in Fig. \ref{fig:limits-dijet}, along the lines discussed in Sec.~\ref{sec:dijet}. The precise regions allowed or ruled out depend on the dominant $Z'$ production mechanism -- and bounds similar to those shown
in  Fig. \ref{fig:limits-dijet}  can be derived separately for each production mode.

\section{Summary of constraints}

The summary of all constraints is presented 
in Figs. \ref{fig:limits}, \ref{fig:limits-general} and Fig. \ref{fig:limits-dijet} in the $\rm log [\zeta^{\mu}]$ vs $M_{Z'}$ and 
$\rm log [\zeta^{j}]$ vs $M_{Z'}$ plane respectively, for the illustrative values of $C^\mu_9$ and $C^\mu_{10}$ chosen. We note that for both the minimal and the general cases, the constraints are similar in nature:

\begin{itemize}

\item  There is a theoretical lower limit on $\zeta^\mu$ imposed 
 by perturbativity requirements on the $Z'$ coupling to fermions; it is described by 
Eq. \ref{eq:zeta-low-perturbative} and Eq. \ref{eq:pertgeneral} and is presented in solid blue lines in the figures. This is a strict lower limit if the theory is to be self-consistent.

\item The narrow width approximation imposes a somewhat softer lower limit on $\zeta^\mu$ if we take $\Gamma/m_{\Zp}\leq 0.1$. We chose to impose this limit to make 
sure that interference efects between BSM physics and the SM 
can be neglected; however the precise definition of what is deemed ``narrow" can potentially vary from the value of 0.1. This limit is presented in pink dashed 
lines in all the plots.

\item In Figs. \ref{fig:limits} and \ref{fig:limits-general}, the region below the light blue dotted line is consistent with having the $Z'$ boson explain the $R_{K}$ anomalies observed by LHCb. 
Combining this with the above two constraints defines the pink shaded region as the region of interesting parameter space for our investigations. 

\item Some important constraints originate from LHC searches for new resonances decaying to  dileptons. Limits on the cross-section for a narrow dimuon resonance imply upper bounds on $\zeta^\mu$ which depend on the partonic production mechanism. These limits are displayed as diagonal lines in Figs. \ref{fig:limits} and \ref{fig:limits-general}. In Fig. \ref{fig:limits}, where only production through $sb$ annihilation is considered, the resulting upper bounds are just beginning to be sensitive to the region which can explain the $R_K$ anomalies. In Fig. \ref{fig:limits-general} we illustrate the dimuon bounds in the cases of production through $b\bar{b}$ and $u\bar{u}$ annihilation as well. In the case of prodution through $b\bar{b}$ annihilation, the constraints are weakened due to the very low $b\bar{b}$ partonic luminosity. In the case of production through $u\bar{u}$ annihilation instead, however, substantial portions of the parameter space which can explain the $R_K$ anomalies are now being probed.

\item In general, as illustrated in Fig.~\ref{fig:limits-dijet} for the case of the minimal model produced through $sb$ annihilation, direct LHC dijet resonance search constraints are weak. Instead, constraints on the dilepton searches, indirectly through the requirements of giving rise the $R_K$ anomalies, can restrict portions of that dijet resonance parameter space.

\item Finally, we illustrate in the general model that, if the number of production modes is increased, then the upper bound on the viable parameter space becomes stronger, because the maximum allowed value of $\zeta^\mu$ is reduced. On the other hand, if the variety of decay modes is increased, then this upper  bound becomes weaker. The perturbativity-derived lower bound  on $\zeta^\mu$ in the general model is always weaker than that in the minimal case; the lower bound corresponding to keeping the $Z^\prime$ narrow is always  tightened.

\end{itemize}

\section{ Conclusion }

In this work we consider the potential of a $\Zp$ boson to explain the recently observed 
$R_{K}$ and $R_{K^{*}}$ anomalies, and to still be consistent with 
latest ATLAS and CMS dijet and dilepton results. To this end, we consider two simple phenomenological
models and work in the language of simplified limits.
We first considered the minimal model needed to explain 
the anomalies, namely a $\Zp$ coupling only to bottom 
and strange quarks and to muons. However
we expect that in a real (UV complete) theory, the $\Zp$
should couple to all the quarks and leptons, and therefore we also considered the constraints in a more general framework. 
For both cases we observe that 
the 13 TeV ATLAS and CMS dilepton results are beginning to constrain certain parts of the paramter space. 
Combining theoretical considerations with
ATLAS and CMS results thus allows us to paint 
a picture on to which any UV complete theory with an extra $\Zp$ model can be mapped. We expect that with the next run of LHCb, 
which can potentially provide further clues to these flavor anomalies, and with the analysis of more ATLAS and CMS data 
on dijet and dilepton resonance searches, the allowed region will be further constrained -- or, possibly, a
new resonance will be discovered. In either case, the language of simplified 
limits can narrow down the interesting region of parameter space consistent with LHC results. 

\section*{Acknowledgments}

This material is based upon work supported, in part, by the National Science Foundation under Grant Nos. PHY-1519045 
(R.S.C., K.M., D.S., and  E.H.S.), and PHY-1417326 (J.I.).  The authors would like to thank Jim Cline, for pointing out a typographic 
error in one of the equations in this manuscript;  our correcting this error yielded significantly stronger constraints on the viable parameter space.

\appendix
\section{Expressions for the decay width}\label{ap1}

In this appendix, we present the expression for the partial decay widths of the $\Zp$(to quarks) used in this paper.  For a $\Zp$ coupled 
to a pair of quarks, we have, 
\begin{align}
\Gamma(\Zp \to  \bar{q^{\prime}} q + \bar{q} q^{\prime} ) =\nu \beta_{\Zp}^2\frac{\left(|c_{L}^{qq^\prime}|^2 + |c_{R}^{qq^\prime}|^2\right)m_{\Zp}}{8 \pi}\ .
\end{align}
Here $\nu =1/2$ if $q=q^\prime$ and $\nu=1$ if $q \ne q^\prime$.

\begin{align}
\Gamma(\Zp \to  \bar{t} q + \bar{q} t ) = \beta_{\Zp}^2\frac{\left(|c_{L}^{tq}|^2 + |c_{R}^{tq}|^2\right)m_{\Zp}}{8 \pi} (1 -x)^2 (2 +x)\ .
\end{align}
Here $x = m_{t}^2 / m_{\Zp}^2$.
\begin{align}
\Gamma(\Zp \to  \bar{t} t ) = \beta_{\Zp}^2\frac{|c_{L}^{tt}|^2 + |c_{R}^{tt}|^2 + \left(|c_{L}^{tt}|^2 + |c_{R}^{tt}|^2 -6c_{L}^{tt}c^{tt}_{R}\right)x}{8 \pi}\cdot m_{\Zp} \sqrt{1 - 4x}\ .
\end{align}
We have neglected mass effects of the top to simplify our discussion. Introducing masses will only reduce the partial width to top final states.
{
Finally for leptons we have
\begin{align}
\Gamma(\Zp \to  \bar{l} l) = \beta_{\Zp}^2\frac{\left(|c_{V}^{\mu}|^2 + |c_{A}^{\mu}|^2\right)m_{\Zp}}{12 \pi}\ .
\end{align}	
}

\bibliographystyle{utphys}

\bibliography{rk_refs} 
\end{document}